\documentclass[reprint, amsmath,amssymb, aps,]{revtex4-1}

\usepackage{graphicx,color}
\usepackage{dcolumn}
\usepackage{bm}

\begin{document}

\preprint{APS/123-QED}

\title{Suppression of magnetism under pressure in FeS: a DFT+DMFT study}


\author{A. V. Ushakov}
\email{ushakov@imp.uran.ru}
\affiliation{M.N. Miheev Institute of Metal Physics of Ural Branch of Russian Academy of
Sciences, 620137, Ekaterinburg, Russia}
\author{A. O. Shorikov}
\affiliation{M.N. Miheev Institute of Metal Physics of Ural Branch of Russian Academy of
Sciences, 620137, Ekaterinburg, Russia}
\affiliation{Ural Federal University, Mira St. 19, 620002 Ekaterinburg, Russia}
\author{N. V. Baranov}
\affiliation{M.N. Miheev Institute of Metal Physics of Ural Branch of Russian Academy of
Sciences, 620137, Ekaterinburg, Russia}
\affiliation{Ural Federal University, Mira St. 19, 620002 Ekaterinburg, Russia}
\author{V. I. Anisimov}
\affiliation{M.N. Miheev Institute of Metal Physics of Ural Branch of Russian Academy of
Sciences, 620137, Ekaterinburg, Russia}
\author{S. V. Streltsov}
\affiliation{M.N. Miheev Institute of Metal Physics of Ural Branch of Russian Academy of
Sciences, 620137, Ekaterinburg, Russia}
\affiliation{Ural Federal University, Mira St. 19, 620002 Ekaterinburg, Russia}

\date{\today}

\begin{abstract}
We investigate the evolution of the magnetic properties in FeS under pressure, and show that these cannot be explained solely in terms of the spin state transition from a high to low spin state due to an increase of the crystal field. Using a combination of density functional theory and dynamical mean field theory (DFT+DMFT), our calculations show that at normal conditions the Fe$^{2+}$ ions are in the $3d^6$ high spin ($S=2$) state, with some admixture of a $3d^7 \underline{L}$ ($S=3/2$) configuration, where  $\underline{L}$ stands for the ligand hole. Suppressing the magnetic moment by uniform compression is related to a substantial increase in electron delocalization and occupation of several lower spin configurations.  The electronic configuration of Fe ions cannot be characterized by a single ionic state, but only by a mixture of the $3d^7 \underline{L}$, $3d^8 \underline{L}^2$, and  $3d^9 \underline{L}^3$ configurations at pressures $\sim$ 7.5 GPa. The local spin-spin correlation function shows well-defined local magnetic moment, corresponding to a large lifetime in the high spin state at normal conditions. Under pressure FeS demonstrates a transition to a mixed state with small lifetimes in each of the spin configurations. 
\end{abstract}


\maketitle

\section{\label{sec:level1} Introduction\protect\\ }
The study of transition metal  compounds with strong electronic correlations is one of the most rapidly developing fields of modern condensed matter physics. This field is mainly connected with the highly unusual physical effects observed in these systems.~\cite{khomskii2014} Most of the research therein has been focused on oxides. However, sulfides, selenides, and chlorides also demonstrate interesting and nontrivial physical phenomena. These include  superconductivity,~\cite{gamble1971} inversion of the crystal field splitting,~\cite{Ushakov2011} multiferroicity~\cite{Singh2009} and many others.

Fe$_{1-x}$S, one of the most widespread sulfides on the Earth (it is also found in cores of terrestrial planets and in lunar and meteoric samples~\cite{Fei-95,Kamimura-92,Kusaba-97,Taylor-70,Wang-06}), is of interest due to its unusual magnetic behavior. The magnetic properties of Fe$_{1-x}$S  change substantially under pressure ($P$), e.g. stoichiometric FeS exhibits three pressure-induced phase transitions.~\cite{Marshall-00} Despite the great importance in understanding these transitions --- not only for condensed matter physics, but also for geoscience and astrophysics --- their mechanism is still unknown. 

At ambient pressure and room temperature FeS crystallizes in the troilite structure (FeS-I phase) with a hexagonal space group $P\bar{6}2c$.~\cite{Marshall-00} In this structure the Fe ions are shifted away from their ideal positions in a NiAs cell, by forming Fe$_3$ triangles in the $ab$ plane.~\cite{Bertaut-56} Troilite transforms to the NiAs structure  above 413~K at normal pressure. The onset of long range magnetic order is observed at $T_N \sim 600$~K (so that FeS in the troilite structure is always magnetically ordered). At normal conditions, the local magnetic moments $\mu \sim$3.2$\mu_B$~\cite{Marshall-00} are aligned ferromagnetically in the $ab$ planes, and are stacked antiferromagnetically along the $c$ axis.~\cite{Andressen1960,Marshall98} At $P=3.4$ GPa troilite transforms to a MnP-type structure (FeS-II) with the orthorhombic space group $Pnma$.~\cite{King-82,Keller-90} The local magnetic moments gradually decrease with pressure, changing with the rate $\sim$0.06$\mu_B/$GPa in FeS-I phase and $\sim$0.08$\mu_B/$GPa in FeS-II phase.~\cite{Marshall-00} An abrupt breaking of long range magnetic order at room temperature is observed above 6.7 GPa.~\cite{King-78,Kobayashi-97,Marshall-00} This transition is accompanied by lattice volume collapse~\cite{Kamimura-92} and a change in the crystal symmetry (space group $P2_1/a$). The effective magnetic moment in Curie-Weiss theory, $\mu_{eff}$, of FeS at ambient pressure is $5.5\mu_B$. At $P=6.7$ GPa (FeS-III phase) $\mu_{eff}$ is $2.2\mu_B$ only.~\cite{Rueff-99} A further increase in pressure leads to a phase transition at $\sim 40$ GPa to the MnP-type nonmagnetic metallic structure (FeS-IV).~\cite{Ono-08} The magnetic transition at 6.7 GPa is the subject of this study.

 There are two possible ways to explain magnetic transitions in Fe$_{1-x}$S sulfides.~\cite{Satpathy-96} The first one is based on the assumption that iron 3$d$ electrons are localized (e.g. due to a Hubbard $U$). At low pressure Fe$^{2+}$ ion has six 3$d$ electrons and is in the high spin (HS) configuration $t_{2g}^{4}e_{g}^{2}$ ($S=2$). With an increase in pressure the crystal field splitting between $t_{2g}$ and $e_g$ sub-shells increases and the low spin (LS) state with $t_{2g}^{6}e_{g}^{0}$ ($S=0$) electronic configuration becomes preferable. Very similar behavior is observed or expected in many other transition metal compounds based on Co$^{3+}$, Fe$^{2+}$, and Fe$^{3+}$ ions.~\cite{korotin1996,Streltsov2011,khomskii2014} However, there is an alternative scenario, which implies gradual metalization under pressure and the loss of the magnetic moments.~\cite{Kobayashi-97,Rozenberg1997} The absence of long range magnetic order in the high-pressure phase can be explained by electron delocalization: the increase of pressure results in the band broadening and in breaking of the Stoner criterion.

It is not clear which model --- based on either localized or itinerant electrons --- is more appropriate for the description of the magnetic properties of Fe sulfides under pressure. In the present work we tackle this problem with first principles calculations using the DFT+DMFT method. We found that at ambient pressure Fe $3d$ electrons can be considered as localized and Fe is predominantly in the $ 3d^6$ high spin state with an admixture of $3d^7 \underline{L}$ configuration.  At higher pressures we observe both a delocalization of the Fe electrons and stabilization of the solution with lower spin. Thus, pressure-induced magnetic transition in FeS is of complex nature and cannot be considered as a pure spin state or as Stoner-like transition.

\section{\label{sec2} Calculation details}
The crystal structures at ambient pressure (AP) and 7.5 GPa (FeS-I, and FeS-III, respectively) were obtained from Ref.~[\onlinecite{Marshall-00}] and  Ref.~[\onlinecite{Nelmes-99}], respectively.
In order to investigate the electronic and magnetic properties of FeS, the  DFT+DMFT  approach~\cite{Anisimov-97} was used. 
This method allows one to treat correlation effects and takes into account realistic band structure, and the first LDA+DMFT calculations on FeS indeed show the importance of correlation effects on the spectral properties.~\cite{Craco2016} The DFT part was obtained using the pseudopotential method as realized in the Quantum Espresso code.~\cite{Gianozzi-09} The exchange-correlation potential was taken in the form proposed by Perdew {\it et al.}~\cite{Perdew-96}
The $k$-grid consisted of 876 points in the whole Brillouin zone, and the wavefunction cutoff was chosen to be $40$ Ry. 

A small, non-interacting, DFT Hamiltonian (H$_{DFT}$) including the Fe $3d$ and S $3p$ states  was generated using the Wannier projection procedure.~\cite{Korotin-08} The unit cell in both phases contains $12$ f.u.
The DFT+DMFT Hamiltonian is written in a form:
\begin{equation}
\label{heis}
 \hat{H} = \hat{H}_{DFT} - \hat{H}_{dc} + \frac{1}{2}\sum_{imm'\sigma\sigma^{\prime}} U_{mm'}^{\sigma\sigma^{\prime}} \hat{n}_{m\sigma,i}\hat{n}_{m'\sigma,i},
\end{equation}
where $U_{mm'}^{\sigma \sigma^{\prime}}$ is the Coulomb interaction matrix, $\hat{n}_{mm',i}^{d}$ is the occupation number operator for the $d$ electrons, $m$ and $m'$ numerates orbitals, while $\sigma$ and $\sigma^{\prime}$ are spins of the electrons on the $i$th site. In order to exclude the $d$-$d$ interaction taken already into account in DFT we used double-counting  correction calculated as $\hat{H}_{dc} = \widetilde{U}(n_{dmft}-1/2)\hat{I}$. Here, $n_{dmft}$ is the self-consistent total number of correlated $d$ electrons obtained within the DFT+DMFT, $\widetilde{U}$ is the average Coulomb parameter for the $d$-shell, $\hat{I}$ is the unit operator. 

\begin{figure}[t!]
\begin{center}
\includegraphics[width=3.5in]{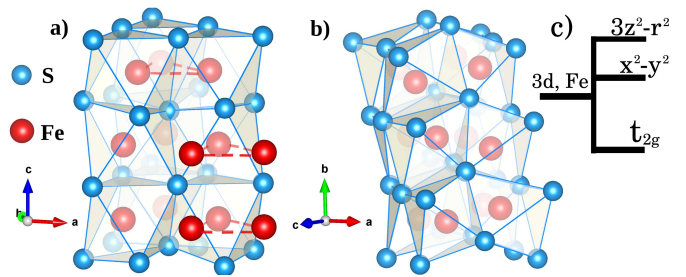}
\end{center}
\caption{\label{struct}(color online) Crystal structure of FeS a) at ambient pressure ($P$-$62c$ space group) and b) at 7.5 GPa ($P2_1$/$c$ space group). c) The schematic crystal field splitting of the Fe $3d$ shell at 0 GPa. The crystal structures were drawn using VESTA.~\cite{VESTA}}
\end{figure}

The DFT+DMFT calculations were performed with the AMULET code~\footnote{$\mathcal{AMULET}$ http://amulet-code.org}, which was previously used to study transition metal compounds such as TiO$_2$, VO$_2$, Li$_2$RuO$_3$ and many others.~\cite{Biermann-05,Katanin-10,Pchelkina-15,Belozerov-12,Shorikov2015,Skorikov2015}
The elements of $U_{mm'}^{\sigma\sigma^{\prime}}$ matrix were parameterized by $U$ and $J_H$ as described in Ref.~[\onlinecite{Liechtenstein-95}]. The values of the Coulomb repulsion parameter $U$ and Hund's coupling constant $J_H$ were taken to be $U = 6$ eV and $J_H = 0.95$ eV.~\cite{Gramsch-03,Ary-06} The impurity solver used in DMFT calculations was based on the segment version of the hybridization expansion Continuous Time Quantum Monte-Carlo method (CT-QMC).~\cite{Werner-06} 
The DFT Hamiltonian was rotated to the local coordinate system, where 
 $d-d$ blocks of $\sum_{\vec k} H_{DFT}(\vec k)$ are diagonal. The experimental antiferromagnetic structure (AFM), with spins in the $ab$ plane ordered ferromagnetically, but stacked antiferromagnetically along the $c$ axis,~\cite{Andressen1960} was used for the ambient pressure phase. 


\section{\label{sec3} CRYSTAL STRUCTURE AND GGA results for ambient pressure}
\begin{figure}[t!]
\begin{center}
\includegraphics[width=3.5in]{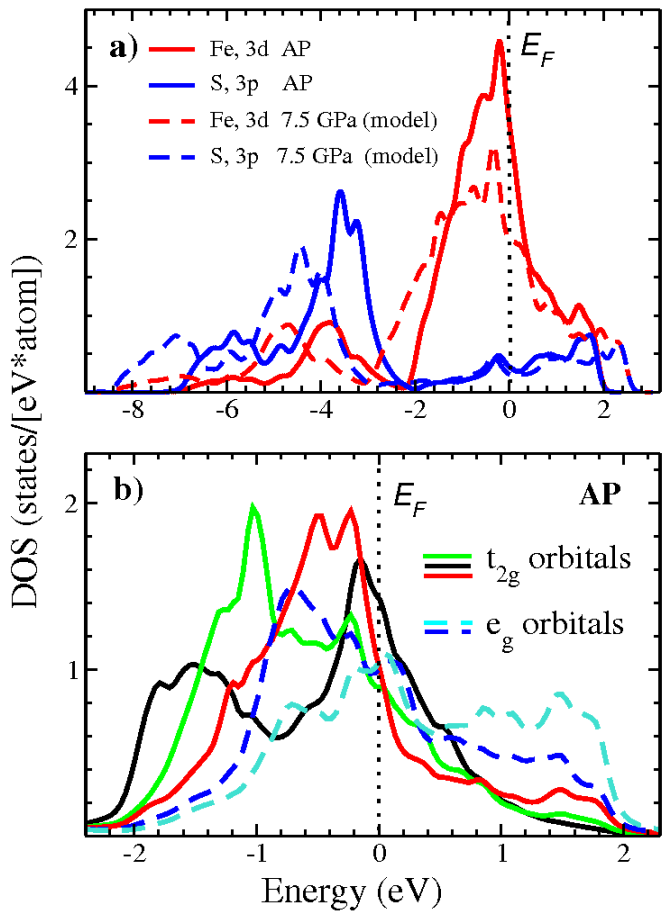}
\end{center}
\caption{\label{dos}(color online) The DFT (nonmagnetic) partial densities of states (DOS) for experimental crystal structure for ambient pressure (AP) and model (artificial) structure for $P=7.5$ GPa (described in Sec.~\ref{sec4}). The local coordinate systems, where $\sum_{\vec k} H_{DFT} (\vec k)$ is diagonal, was used. The Fermi level is in zero.}
\end{figure}

We start with a discussion of the nonmagnetic DFT results at AP. The density of states (DOS) is shown in Fig.~\ref{dos}a. The S $3p$ states are placed approximately from -7 to -2 eV, while Fe $3d$ orbitals are in the vicinity of the Fermi level. 

Fig.~\ref{dos}b illustrates the contributions from different $3d$ orbitals to the total density of states. The FeS$_6$ octahedra are strongly distorted in the low pressure phase. There are three short (2.36, 2.38, and 2.42 \AA), two intermediate (2.51 and 2.56 \AA) and one long (2.72 \AA) Fe-S bonds. This pyramidal-like surrounding (with one very long Fe-S bond) results in a strong splitting of $e_g$ levels. One might expect that in this situation the $x^2-y^2$ orbital would go higher in energy than $3z^2-r^2$ orbital (here and below, unless stated otherwise, the local coordinate system with axes directed towards S ions is used). This is, however, not the case in FeS, since Fe is shifted (by $\sim$0.3 \AA) inside of the FeS$_5$ pyramid, which weakens hybridization between S $3p$ and Fe $x^2-y^2$ orbitals considerably. As a result, $3z^2-r^2$ orbital turns out to be roughly $0.3$ eV higher than the $x^2-y^2$ orbital (as determined by diagonalizing the on-site Hamiltonian $H_{DFT}$). The $t_{2g}$ orbitals of Fe are nearly degenerate.
\begin{figure}[t!]
\begin{center}
\includegraphics[width=3.5in]{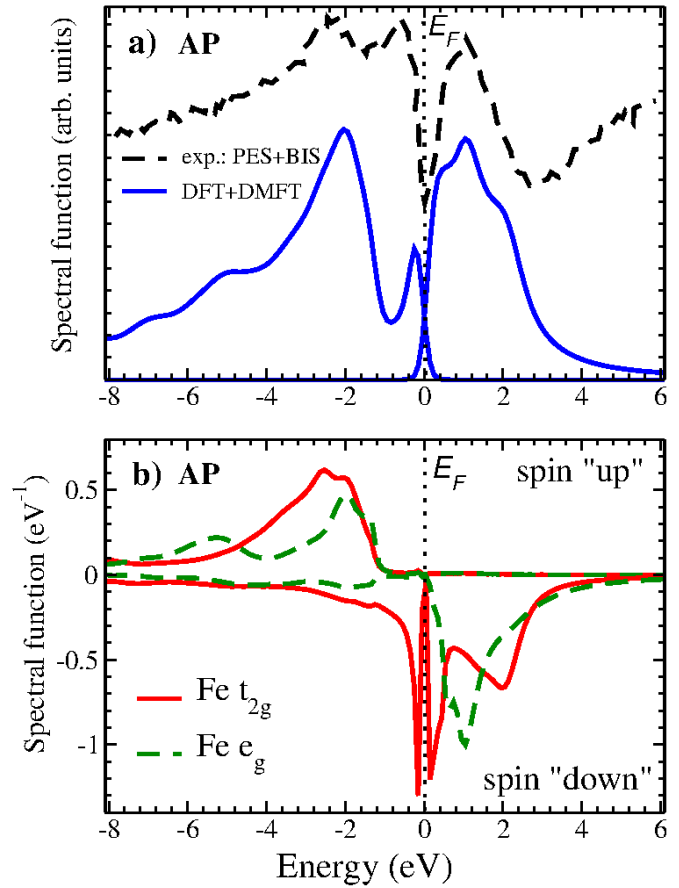}
\end{center}
\caption{\label{specfunc}(a) Experimental photoemission and inverse photoemission spectra\cite{Shimada-98} together with the DFT+DMFT spectral function (for the AFM solution) as obtained by the Pad\'e approximation. Theoretical spectral functions were weighted by photoionization cross-sections\cite{robinson1991} and broadened by 0.2 eV. (b) Spin- and orbital-resolved spectral functions for ambient pressure (AP). All graphs correspond to inverse temperature $\beta=1/T$=$30$ eV$^{-1}$.} 
\end{figure}

Another quite important effect is related to the fact that three Fe ions form isolated triangles, shown by dashed lines in Fig.~\ref{struct}a, sharing edges of the FeS$_6$ octahedra (there are two short, 2$\times$2.92 \AA, and four long, 2$\times$3.67 \AA ~and 2$\times$3.80 \AA, Fe-Fe bonds). As a result there are $t_{2g}$ orbitals on different Fe sites, which are directed towards each other. This leads to a strong direct overlap and a bonding-antibonding splitting for two out of three $t_{2g}$ orbitals, which is $\sim$1.3 eV for one of the $t_{2g}$ orbitals. This is clearly seen in the partial DOS as a two peak structure (green and black curves in Fig.~\ref{dos}b). This interpretation agrees with the estimates of the hopping parameter for the $xy$ orbitals (black) of 0.5-0.6 eV. Similar features of the electronic structure were found in many other systems with the edge sharing geometry,~\cite{Streltsov2014,khomskii2014} and it was shown that they may strongly affect the magnetic properties of the system.~\cite{Streltsov2012a,Streltsov2016a}

Both these effects: direct metal-metal bonding and, especially, the strong distortion of the Fe-ligand octahedra, result in the situation whereby all five $d$ orbitals appear in more or less the same energy interval. A naive ionic model, based on the competition between the $t_{2g}-e_g$ crystal field splitting and Hund's rule coupling, $J_H$,~\cite{khomskii2014} does not work here. This is in strong contrast with many other materials, where such a simplified treatment provides a very good description of the spin-state transitions.~\cite{khomskii2014,fazekas}
Hence, there is no other way to describe pressure-induced magnetic transition in FeS than a direct calculation, which takes into account both the peculiarities of the crystal structure and presence of the strong Coulomb interaction, i.e. Hund's exchange $J_H$ and Hubbard $U$. 

\section{\label{sec4} DFT+DMFT RESULTS AND DISCUSSION \protect\\ }
\subsection{Ambient pressure phase}
The DFT+DMFT calculations of the AFM AP phase show that FeS is an insulator with a tiny band gap of $\sim$20 meV.~\footnote{In order to calculate spectral functions we used the Pad approximation for the self-energy $\Sigma (i \omega_n)$, which results in some uncertainty in estimation of the band gap.} This agrees with an experimental estimate of the activation energy (40 meV).~\cite{Gosselin1976} 

A comparison between the experimental and theoretical spectral functions is shown in Fig.~\ref{specfunc}. The DFT+DMFT spectral function reproduces all of the main features of the experimental spectra. The peak at $\sim -2.4$ eV  in the photoemission data originates from the Fe $t_{2g}^{\uparrow}$ and $e_{g}^{\uparrow}$ states, while the one at $\sim -0.6$ eV corresponds to a singly occupied $t_{2g}^{\downarrow}$ ($xy^{\downarrow}$) orbital. 
Thus, analysis of the spectral functions shows that Fe ions are in the high spin state at AP. Moreover, this result suggests that one may study the pressure-induced magnetic transition in Fe$_{1-x}$S or Fe$_{1-x}$Se by tracking this feature at $\sim -0.6$ eV in photoemission measurements. The peak at $\sim 1$ eV and the shoulder at $\sim 0.4$ eV in the BIS spectra correspond to the $e_{g}^{\downarrow}$ and $t_{2g}^{\downarrow}$ states.

\begin{figure}[t!]
\begin{center}
\includegraphics[width=3.5in]{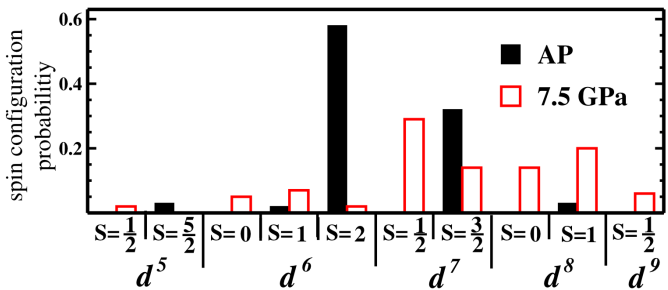}
\end{center}
\caption{\label{dia}(color online) Weights (i.e. statistical probability), $w_i$, of various electronic configurations at AP and at 7.5 GPa phases calculated by the DFT+DMFT method for paramagnetic state,  $\sum_i w_i = 1$. For the high-pressure phase averaged data over three crystallographically inequivalent Fe ions are presented. The number of $t_{2g}$ ($e_g$) electrons is 4.09 (2.38) at AP. In the high-pressure phase there are three different Fe, corresponding $t_{2g}$ ($e_g$) occupancies  are 5.14, 4.73, and 4.42 (2.19, 2.59, and 2.54).
}
\end{figure}

The DFT+DMFT calculations were carried out for both the experimental AFM structure and the standard paramagnetic (PM) regime, where self-energy is averaged over spins. If this type of averaging is not included, the calculation converges to a magnetic solution. The fragile insulating state does not survive the transition to paramagnetic state. The average magnetic moment at ambient pressure $\sqrt{ \langle m_z^2 \rangle}$ is roughly the same in both AFM and PM states:
 $\sqrt{\langle m_z(PM)^2 \rangle}=3.7 \mu_B$ and $\sqrt{ \langle m_z(AFM)^2 \rangle}=3.6 \mu_B$. One can see that these values are close to what one may expect for Fe$^{2+}$ ($d^6$) in the high spin state ($S=2$). Analysis of the different electronic configurations measured in the DMFT solver shows that this a state has the highest probability in our calculations, see Fig.~\ref{dia}. Moreover, the second state having substantial weight, $d^7$, is essentially the same $d^6$ high spin state with one ligand electron added ($t_{2g}^4e_{g}^3$, $S=3/2$). This state can be denoted as $d^7{\underline L}$, where $\underline {L}$ stands for a ligand hole.

It is worth mentioning that at normal conditions FeS orders magnetically and Fe ions essentially adopt a single ionic configuration ($d^6$ high spin with a single electron in minority spin sitting on the $xy$ orbital), which explains why DFT calculations based on the local density approximation + Hubbard $U$ (LDA+U) successfully reproduce the band gap and experimental photoemission spectra.~\cite{Rohrbach2003} We will show later that at higher pressure the Fe ion is in mixture of different electronic configurations. This type of state obviously cannot be described by any method based on single-determinant wavefunctions (such as LDA+U).

The exchange constants for the Heisenberg model in spin-polarized DFT+DMFT, i.e. in an AFM solution, can be calculated as:~\cite{Katsnelson2000,Secchi2014}
\begin{eqnarray}
J_{ij} = \frac T4  Tr \left(
\Delta^{m}_i (i\omega_n) G^{mm'}_{ij,\uparrow} (i\omega_n)
\Delta^{m'}_j (i\omega_n) G^{m'm}_{ji,\downarrow} (i\omega_n) \right), \nonumber
\end{eqnarray}
where $T$ is the temperature, $i,j$ are the site indexes, the trace is taken over orbital indexes $m,m'$ and Matsubara frequencies $i\omega_n$, $G_{ij}$ is the corresponding inter-site Green function, and $\Delta_i^{m} (i\omega_n)$ is the on-site 
exchange splitting defined only by the electron self-energy $
\Delta_i^m (i\omega_n)  = \Sigma^m_{i,\uparrow} (i\omega_n) - \Sigma^m_{i,\downarrow} (i\omega_n)$. 

We found that in the AP phase the largest exchange constant is $J_c=24$ K (AFM), along the $c$ axis. The exchange coupling between the twelve next nearest neighbors lying in adjacent $ab$ planes is $J_{nn} \sim$ 18 K (AFM). In contrast, the in-plane interaction is rather weak and ferromagnetic, $J_{ab}\sim -1$K. This explains the experimentally observed magnetic structure (ferromagnetic $ab$ plane stacked antiferromagnetically along $c$ axis\cite{Andressen1960,Marshall98}). The exchange parameters in the systems with localized magnetic moments may strongly depend on the Hubbard $U$ (in the simplest approximation $J \sim 2t^2/U$~\cite{khomskii2014}). However, we would like to mention that the present choice of $U$ and $J_H$ gives the Curie-Weiss temperature $\theta =1040$ K (recalculated using these exchange constants), which is quite close to the experimental estimate of $\theta_{exp}=1160$~K.~\cite{Horwood1976}

\subsection{High-pressure phase: P=7.5 GPa}

There are three inequivalent Fe ions in the high-pressure structure (denoted as Fe1, Fe2, and Fe3 throughout the text). Analysis of the magnetic properties shows that local magnetic moment drops down to $\sqrt{ \langle m_z^2 \rangle } = 1.9\mu_B$ (averaged over three inequivalent Fe sites) for $P=7.5$ GPa, which agrees with the experimentally observed suppression of magnetism in FeS.~\cite{Rueff-99,Ono-08} The decrease in the magnetic moment is accompanied by a more uniform probability distribution of different electronic configurations and by an increase in the weights of the states with smaller total spin. Note that the $d^7$ and $d^8$ configurations now have a larger probability, which is related to the increased Fe $3d$ - S $3p$ hybridization due to the decrease of the Fe-S bond distance with pressure (this corresponds to an increase of the weights of the $d^7\underline{L}$ and $d^8\underline{L}^2$ configurations). 
The value of the local magnetic moments $\sqrt{ \langle m_z^2 \rangle }$ for each class of Fe correlates with the mean Fe-S bond distance in the FeS$_6$ octahedra.
The lowest value $\sqrt{ \langle m_{z,Fe3}^2 \rangle }=1.7\mu_B$ corresponds to the average Fe-S bond distance $d_{\langle Fe3-S \rangle}=2.31$\AA, while for Fe2 and Fe1: $d_{\langle Fe2-S \rangle}=2.37$\AA, $d_{\langle Fe1-S \rangle}=2.42$\AA~and $\sqrt{ \langle m_{z,Fe2}^2 \rangle }=1.8\mu_B$, $\sqrt{ \langle m_{z,Fe1}^2 \rangle }=2.2\mu_B$. This is related with an increase of the $t_{2g}-e_{g}$ crystal field splitting with decrease of the Fe-S bond length, which stimulates the spin-state transition.

\begin{figure}[t!]
\begin{center}
\includegraphics[width=3.5in]{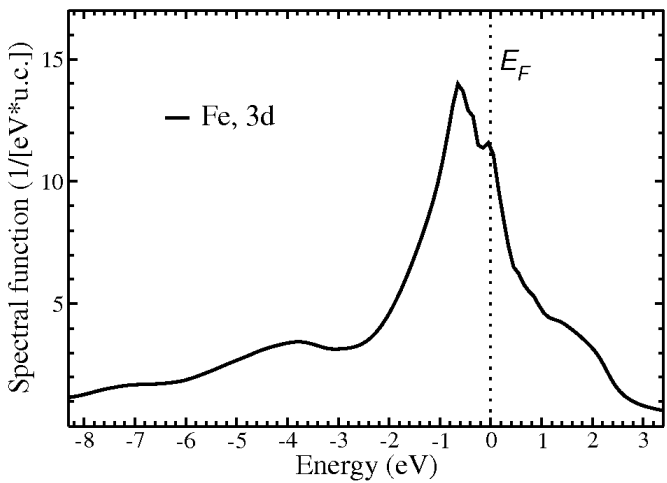}
\end{center}
\caption{\label{dos75}(color online) Spectral function of Fe $3d$ in FeS at 7.5 GPa and $\beta=1/30$ eV$^{-1}$.
}
\end{figure}

As discussed in Sec.~\ref{sec:level1}, the magnetic properties of Fe$_{1-x}$S are usually described based on either localized or itinerant electron models. The former implies that the electrons are localized on atomic sites, and that by increasing pressure we gradually transfer the system from the high spin to low spin state. In the itinerant electron theory one may explain magnetic properties of FeS by gradual metalization and loss of the magnetic moments. The key characteristics, which can discriminate between these two scenarios, is the degree of the spin localization. In DMFT the space-time spin correlators are not easily accessible, but one can calculate the local (in real space) spin-spin correlation function $\langle \hat{S}_z(0) \hat{S}_z(\tau) \rangle$ in the imaginary time ($\tau$) domain. If magnetic moments are localized, this correlator is constant:~\cite{Georges-96}
\begin{equation}
\langle \hat{S}_z(0) \hat{S}_z(\tau) \rangle \sim S_z^2.
\end{equation}
This gives the Curie-Weiss law, if one would recalculate local magnetic susceptibility $\chi_{loc} = \mu_B^2 \int\limits_0^{\beta} \langle \hat{S}_z(0) \hat{S}_z(\tau) \rangle d \tau$ with this substitution. In contrast, an imaginary time dependence of this correlator indicates the delocalization of the spin moments. E.g., in a Fermi-liquid:
\begin{equation}
\label{correl}
\langle \hat{S}_z(0) \hat{S}_z(\tau) \rangle \sim \frac{T^2}{\sin (\tau \pi T)^2}
\end{equation}
for $\tau$ sufficiently far from $0$ and $\beta$.~\cite{Werner2008}
\begin{figure}[t]
\begin{center}
\includegraphics[width=3.5in]{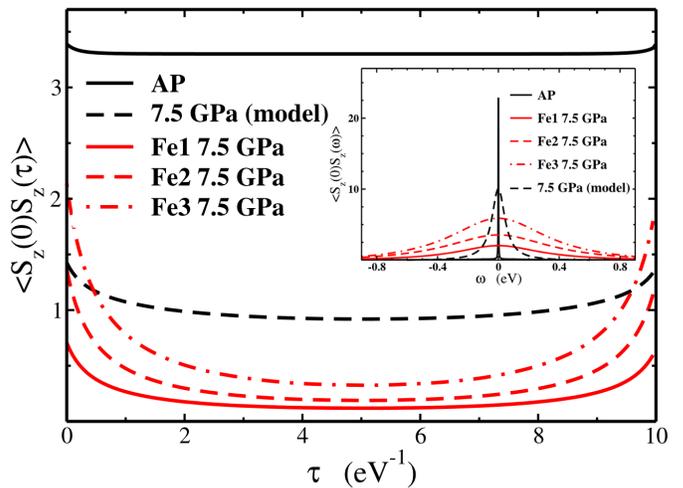}
\end{center}
\caption{\label{corr}(color online) Imaginary time, $\tau$, and frequency, $\omega$, (inset) dependence of the local spin correlator calculated by DFT+DMFT at ambient pressure (AP) and 7.5 GPa for $\beta=10$~eV$^{-1}$ ($T \sim 1100$ K). The black dashed curve corresponds to the calculated model structure as described in Sec.~\ref{sec4}. There is one inequivalent Fe  in both the AP phase and the model structure, while there are three inequivalent Fe for $P=7.5$ GPa.}
\end{figure}

In Fig.~\ref{corr} we show how $\langle \hat{S}_z(0) \hat{S}_z(\tau) \rangle$ behaves for different pressures in FeS. One may see that the spin correlator for ambient pressure is much higher than for $P=7.5$ GPa, which reflects the larger local magnetic moment in the former case.


Secondly, $\langle \hat{S}_z(0) \hat{S}_z(\tau) \rangle$ is almost $\tau-$independent at AP, while for the higher pressure a clear imaginary time dependence is observed, which gives some evidence for spin delocalization at higher pressures.  
It is clearer to consider the dependence of this correlator on real frequencies ($\omega$), as shown in the inset in Fig.~\ref{corr}. Spin excitations at ambient pressure occur in a very narrow frequency range, which corresponds to a large lifetime (the width of the peak is  inversely proportional to the lifetime) and to the local nature of the spin. In contrast, there is substantial dispersion for all inequivalent Fe sites at higher pressure. This demonstrates a change in the behavior of the electrons and suggests a transition to an itinerant regime.  This means that it is incorrect to use oversimplified models like spin-state transition to explain pressure dependence of the magnetic properties in FeS.
\begin{figure}[t!]
\begin{center}
\includegraphics[width=3.5in]{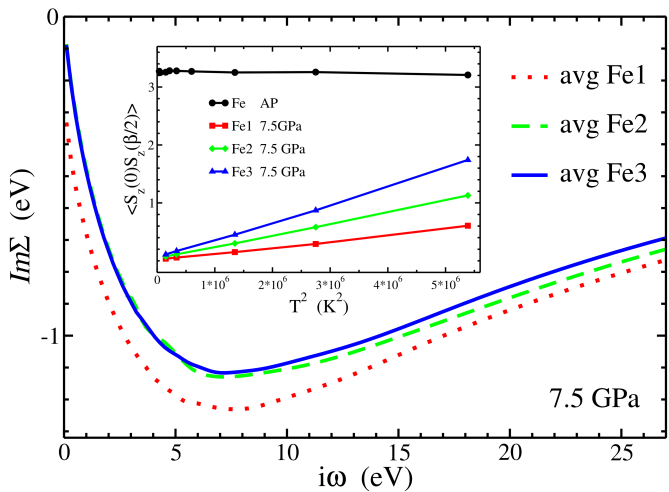}
\end{center}
\caption{\label{sigma}(color online) The dependence of the imaginary part of the self-energy $\Sigma$ on Matsubara frequencies ($i\omega_n$) for $\beta=30$ eV$^{-1}$ at 7.5 GPa. The average curves for 5 different $3d$ orbitals of three Fe impurities are presented, since there is only minimal differences between $\Sigma(i \omega_n)$ for different orbitals. The $T^{2}$ dependence of the $\langle \hat{S}_z(0) \hat{S}_z(\tau) \rangle$ correlator at $\tau=\beta$/$2$ at ambient pressure and 7.5 GPa is shown in inset.}
\end{figure}

The same conclusion can be drawn from analysis of the temperature dependence of the local spin-spin correlator. One may expect that this dependence will be strong in case of a metal. As follows from Eq.~\eqref{correl} in the simplest case of the Fermi-liquid this correlator at $\tau = \beta/2$ is proportional to $T^2$. In contrast, one would not expect any temperature dependence of the correlator in the system with local magnetic moments. In inset of the Fig.~\ref{sigma} we show that in FeS $\langle \hat{S}_z(0) \hat{S}_z(\beta/2) \rangle$ is strictly linear as function of $T^2$ at high pressure and is essentially independent of temperature at normal conditions, wherein the magnetic moments are well-formed and basically frozen at any temperature. 

It is important to stress that, while there is a clear delocalization of the spin moments under pressure, FeS cannot be described in terms of conventional band magnetism, e.g. in the  framework of Stoner theory. Analysis of the electronic configuration distribution presented in Fig.~\ref{dia} shows that there are strong dynamical fluctuations between several ionic configurations in this phase. Such behaviour cannot be described by static mean-field methods like LDA or LDA+U and the use of DMFT is essential. 

Because of these fluctuations, FeS turns out to be metallic in our LDA+DMFT calculations for the high-pressure crystal structure corresponding to $P=7.5$ GPa (see Fig.~\ref{dos75}), while further decreasing the temperature (below $T=$390 K, which was used in our calculation) or small variations the crystal structure may open a band gap.~\footnote{It is rather illustrative that LDA+U calculations with the same $U$ and $J_H$ also give metallic state for FM and AFM solutions, even suppression of dynamic fluctuation does not lead to insulating ground state. This suggests that the crystal structure for $P=7.5$ GPa may be different from what was published previously and used in the present calculations.} It was claimed in Ref.~\cite{Kobayashi2001} that a tiny gap develops in FeS above 6 GPa according to their resistivity measurements. However, one has to be very careful with this interpretation, since there is only a slight and, atypical for semiconductors, decrease in their measured resistivity with temperature. Situations such as these are often observed in transition metal compounds. For example, detailed analysis of the optical data in CaCrO$_3$ revealed that it is a metal, while resistivity shows ``insulating'' behavior (i.e. $\rho(T)$ decreases with temperature)~\cite{Komarek2008}. In any case we do not expect that the presence or absence of a tiny band gap would strongly affect magnetic properties of FeS.
Moreover, the imaginary part of the self-energy, $\Sigma(i \omega_n)$ in our calculations, is practically linear at low frequencies (as shown in Fig. \ref{sigma}) at high pressure. The effective masses extracted from the low-frequency behavior of $Im\Sigma(i \omega_n)$ are $m^*/m \sim 1.6-2.2$ (depending on which inequivalent Fe). 

Several very important structural characteristics, which may trigger the magnetic transition, are changed simultaneously upon applying pressure. First of all, a decrease in the Fe-S bond distance will result in an increase in the $t_{2g}-e_{g}$ crystal field splitting (we have seen, however, in Sec.~\ref{sec3} that this parameter is ill defined in FeS) and therefore may induce the spin-state transition. In turn, a decrease in the Fe-Fe bond distance enhances band dispersion and hence electrons become more itinerant. In addition to these two factors the symmetry of the crystal changes under the pressure, which leads to the destruction of the Fe$_3$ triangles. 

 In order to understand which factor is more important and leads to the magnetic transition, we carried out the DFT+DMFT calculation for a model crystal structure, with the same symmetry as the AP structure, but with unit cell volume corresponding to that at $P=7.5$ GPa. This model calculation results in a $\sim 30 \%$ increase of the Fe $3d$ band width, as one can see from Fig.~\ref{dos}(a). Surprisingly, even this simplified model still describes magnetic transition reasonably well giving $\sqrt{ \langle m_z^2 \rangle} = 2.0 \mu_B$. Thus, the decrease of the volume explains $\sim$ 80\% of the reduction of the magnetic moment in FeS, and a change of the crystal symmetry is not so important for the absolute value of the moment. This mechanism seems to be also responsible for the disappearance of a long-range ferrimagnetic order in pyrrhotite-type compounds (Fe$_{1-y}$Co$_y$)$_7$X$_8$ (X = S, Se) with increasing Co concentration.~\cite{Sato-85, Sato-92, Baranov-15}
 
However, while uniform compression can basically explain the suppression of the local magnetic moment, it reproduces increase of spin delocalization  only partially. This is clearly seen from Fig.~\ref{corr}; the change in volume is not enough to sufficiently reduce the local spin-spin correlator. Moreover, it is the change of the crystal symmetry which makes the Fe ions so different from the point of view of spin delocalization (the $\tau$ dependence is very different for these three inequivalent Fe).

\section{Conclusion}
We found that due to a strongly distorted crystal structure the $t_{2g}$ and $e_g$ bands are not well separated in FeS, so that one may hardly define the $t_{2g}-e_g$ crystal field splitting. At ambient pressure the Fe ions are in the high spin $d^6$ state with some admixture of the $d^7 \underline {L}$ state due to a large covalency. Under pressure we observe (1) suppression of the spin moment and (2) strong dynamical fluctuations between different ionic configurations, which cannot be described by the static mean-field methods like LDA+U. The spin moment is largely suppressed, but still non-zero at $P=7.5$ GPa ($\sqrt {\langle m_z^2 \rangle}=1.9 \mu_B$).  This resembles a pressure-induced spin-state transition, driven by increase of the $t_{2g}-e_g$ crystal field splitting. However, simultaneously with the suppression of the magnetic moment, the frequency dependence of the local spin-spin correlator shows that the Fe $3d$ electrons become delocalized. Thus, we show that one cannot describe the pressure-induced magnetic transition in FeS as a standard spin-state transition. 

The changes in the magnetic properties of FeS with pressure are related to a modification of the electronic structure. In the high-pressure phase FeS is a metal or close to metallic regime, effective masses $m^*/m \sim 1.6-2.2$, while at normal conditions FeS is an insulator with well defined local magnetic moments. This study opens new perspectives for investigation of the magnetic properties not only of FeS, but also of other Fe and Co sulfides and selenides, where very similar magnetic transitions were observed.~\cite{Shimada-99, Hobbs-99, Wang-06, Sato-88, Liebsch-10, Kamimura-92}

\section*{Acknowledgments} We are grateful to E. Kurmaev for various stimulating discussions on the electronic and magnetic properties of FeS and to J. McLeod for critical comments to the text of the paper. This work was supported by the grant of the Russian Scientific Foundation (project no. 14-22-00004). 

\bibliography{./1}

\end{document}